\documentclass[manuscript]{aastex61}
\usepackage{bm}

\received{2017 December 27}
\revised{2018 February 18}
\accepted{2018 February 28}

\shorttitle{Particle Number Dependence of Moon Forming Simulation}
\shortauthors{Sasaki and Hosono}

\begin{document}

\title{Particle Number Dependence of The $N$-Body Simulations of Moon Formation}

\correspondingauthor{Takanori Sasaki}
\email{takanori@kusastro.kyoto-u.ac.jp}

\author[0000-0003-1242-7290]{Takanori Sasaki}
\affil{Department of Astronomy, Kyoto University, Kitashirakawa-Oiwake-cho, Sakyo-ku, Kyoto 606-8502, Japan}

\author{Natsuki Hosono}
\affiliation{Japan Agency for Marine-Earth Science and Technology, 2-15, Natsushima-cho, Yokosuka-city, Kanagawa 237-0061, Japan}
\affiliation{RIKEN Advanced Institute for Computational Science, 7-1-26 Minatojima-minami-machi, Chuo-ku, Kobe, Hyogo, Japan}

\begin{abstract}

The formation of the Moon from the circumterrestrial disk has been investigated by using $N$-body simulations with the number $N$ of particles limited from $10^4$ to $10^5$. We develop an $N$-body simulation code on multiple Pezy-SC processors and deploy FDPS (Framework for Developing Particle Simulators) to deal with large number of particles. We execute several high- and extra-high-resolution $N$-body simulations of lunar accretion from a circumterrestrial disk of debris generated by a giant impact on Earth. The number of particles is up to $10^7$, in which 1 particle corresponds to a 10 km-size satellitesimal. We find that the spiral structures inside the Roche limit radius differ between low-resolution simulations ($N \leq10^5$) and high-resolution simulations ($N \geq10^6$). According to this difference, angular momentum fluxes, which determine the accretion timescale of the Moon also depend on the numerical resolution.

\end{abstract}


\section{Introduction}

The giant impact hypothesis is a widely accepted scenario for the origin of the Moon (Hartmann \& Davis 1975; Cameron \& Ward 1976) and has been favored because it can explain the Moon's mass and iron depletion, as well as the angular momentum of the Earth-Moon system (Canup \& Asphaug 2001). The giant impact process has been investigated by smoothed particle hydrodynamics (SPH) simulations (see, e.g., Benz et al. 1986; Canup 2004), which showed that such giant impacts usually result in the formation of a circumterrestrial disk whose mass is usually smaller than 2.5 $M_L$, where $M_L$ is the present lunar mass ($\sim 0.0123 M_{\oplus}$, where $M_{\oplus}$ is the present Earth mass). Most of the disk material is distributed near or inside the Roche limit radius $R_\mathrm{Roche}$ ($\sim 2.9 R_{\oplus}$, where $R_{\oplus}$ is the present Earth radius), if the orbital angular momentum of the impactor is about the same as that of the present Earth-Moon system.

Lunar accretion from the circumterrestrial disk was investigated by using $N$-body simulations (Ida et al. 1997; Kokubo et al. 2000). Ida et al. (1997) modeled the circumterrestrial disk by 1000-1500 particles. However, this value is insufficient to resolve the spatial structure of the disk, which is important for angular momentum transfer and mass transfer in the disk. Kokubo et al. (2000) performed higher-resolution $N$-body simulations by using 10000 particles for the disk and found that the results coincided qualitatively and quantitatively with that of Ida et al. (1997) for the same surface density of the disk. To more clearly see the evolution of the spatial structure of the disk, Kokubo et al. (2000) also performed an $N$-body simulation based on the ``rubble pile model," in which no mergers were allowed. To date, this 10000-particle $N$-body simulation is the only rubble pile model simulation to follow the process of lunar accretion until Moon formation.

Conversely, Takeda \& Ida (2001) evaluated the angular momentum transfer processes in the circumterrestrial disks, as they followed the lunar accretion process halfway. They performed global $N$-body simulations based on the rubble pile model with particle numbers up to $10^5$ to represent the disk. They found that angular momentum transfer was dominated by the gravitational torque due to the spiral structure formed within the Roche limit radius around the Earth and by the collective motion of the particles associated with the structure. They concluded that the formation and evolution of the spiral structure were regulated only by the disk surface density such that the angular momentum transfer depends on only the surface density but not on the particle number $N$ when $N \geq 10^3$.

However, note that no simulations exist that focus on how the satellite mass and growth speed depend on the number of particles. At the very first stage of the accretion of the Moon, the radius of debris particles is thought to be approximately 10 km, which corresponds to numerical simulations with approximately $10^7$ particles.

To address this shortcoming, we present herein the results of $N$-body simulations based on the rubble pile model with $N$ = $10^4$, $10^5$, $10^6$, and $10^7$ particles, which allow us to follow the lunar accretion process until Moon formation. We investigate how the numerical resolution affects the formation and evolution of the disk structure, including the spiral structure. We also compare the results to those of Kokubo et al. (2000) and Takeda \& Ida (2001). We find that the spiral structures inside the Roche limit radius are the same as those found in previous studies based on low-resolution simulations ($N \leq 10^5$) but differ from those found in high-resolution simulations ($N \geq 10^6$). On the basis of these differences, we find that the angular momentum fluxes, which determine the accretion timescales of the Moon, also depend on the numerical resolution of the simulation.

We assume herein that the circumterrestrial disk is an entirely particulate disk. SPH simulations suggest that most particles are in vapor or liquid phase just after the giant impact, with an average temperature exceeding 4000 K (Canup 2004). In such cases, lunar accretion proceeds on the cooling timescale of the entire disk (see, e.g., Thompson \& Stevenson 1988; Ward 2012) or by the slow spreading of the inner melt disk (Salmon \& Canup 2012), both of which are much longer ($\sim$ 100 yr) than the accretion timescale predicted by $N$-body simulations ($\sim$ 1 month). While the disk's thermal evolution would depend on how the disk made of a vapor-liquid-solid mixture cools and spread, the resulting mass of the Moon is largely determined by the initial mass and angular momentum of the disk produced by the giant impact rather than the subsequent disk evolution (Salmon \& Canup 2012). The purpose of the present study is to evaluate how the lunar accretion process depends on the number $N$ of $N$-body simulations. Therefore, we do not pursue such long-term disk evolution and consider only the same situations adapted in previous studies (Ida et al. 1997; Kokubo et al. 2000; Takeda \& Ida 2001).

Section 2 describes the model and calculation method. Section 3 presents the results of the $N$-body simulations. Section 4 is devoted to a summary and some discussions.

\section{Method of Calculation}

We performed several sets of high- and extra-high-resolution $N$-body simulations that included physical collisions and self-gravity to simulate lunar accretion from a circumterrestrial debris disk generated by a giant impact on Earth. To investigate how the lunar accretion process depends on the number $N$ of $N$-body simulations, we adopt physical and numerical models similar to those used in previous studies (Ida et al. 1997; Kokubo et al. 2000), except for the number of particles.

We start the simulations of lunar accretion by assuming a solid particle disk, which forms through the condensation of the silicate vapor-liquid generated by the giant impact (Canup 2004). We set the density $\rho_{\oplus}$ of the proto-Earth to 5.5 g cm$^{-3}$ (equal to the present bulk density of Earth) and the density $\rho_p$ of the disk particles to 3.3 g cm$^{-3}$ (equal to the present bulk density of the Moon). We scale the orbital radii by the Roche limit radius as follows:
\begin{equation}
R_\mathrm{Roche} = 2.456 \left( \frac{\rho_p}{\rho_{\oplus}} \right)^{-1/3} R_{\oplus} \simeq 2.9 R_{\oplus} ,
\end{equation}
where $R_{\oplus}$ is the Earth radius. We use the proto-Earth mass as the present Earth mass $M_{\oplus} = 5.97 \times 10^{27}$ g, and we assume a disk of equal-mass particles. The physical radius of disk particles with mass $m$ is
\begin{equation}
r_p = \frac{1}{2.456} \left( \frac{m}{M_{\oplus}} \right)^{1/3} R_\mathrm{Roche},
\end{equation}
following Kokubo et al. (2000).

The orbits of disk particles are calculated by numerically integrating the equation of motion,
\begin{equation}
\frac{d \bm{v}_i}{dt} = -G M_{\oplus} \frac{\bm{x}_i}{|\bm{x}_i|^3} - \sum^N_{j \neq i} Gm_i \frac{\bm{x_i}-\bm{x}_j}{|\bm{x}_i-\bm{x}_j|},
\end{equation}
where $\bm{x}$ and $\bm{v}$ are the position and velocity of disk particles, respectively, and $G$ is the gravitational constant. For numerical integration, we follow Kokubo et al. (2000) and use a modified fourth-ordered Hermite scheme for planetary $N$-body simulation with a shared timestep of $2^{-9}$. A full description of the scheme is given in Makino \& Aarseth (1992).

We adopt the rubble pile model for physical collisions in which no mergers are allowed and instead particles simply bounce inelastically off each other. In this model, gravitationally bound aggregates form outside the Roche limit radius, whereas aggregates are tidally disrupted when they stray too close to the proto-Earth. For simplicity, we do not consider fragmentation, heat generation, or vaporization. We assume that the central body is spherical and do not include the tidal force due to tidal bulges raised on Earth, because the timescale of the tidal orbital evolution is much longer than the accretion timescale. When particles fall inside the central body's radius $R_{\oplus}$, we assume they are accreted onto the Earth and remove them from the calculation.

In addition to self-gravity, we must resolve collisions between particles. Toward this end, Ida et al. (1997) assumed that when two particles overlap, the particles always and immediately merge into a single particle. Kokubo et al. (2000) allowed the merger of two particles only if the Jacobi energy of the two particles was negative after the collision; otherwise, they rebounded. Note that merging particles would lead to an overestimation of the final satellite mass because the mass loss due to tidal stripping and/or collisions between rabble piles is ignored. Thus, we do not allow particle mergers in the present work; instead, all collisions between two particles are resolved by rebound.

When the distance between two particles is less than the sum of the particle radii, we change the particle velocities according to the restitution coefficient $\epsilon$ ($0 \leq \epsilon \leq 1$),
\begin{eqnarray}
\bm{v'}_\mathrm{n} &=& -\epsilon_\mathrm{n} \bm{v}_\mathrm{n}\\
\bm{v'}_\mathrm{t} &=& \epsilon_\mathrm{t} \bm{v}_\mathrm{t},
\end{eqnarray}
where $\bm{v}$ and $\bm{v'}$ are the relative impact velocity and the relative rebound velocity, respectively; subscripts n and t indicate the normal and tangential components, respectively. We assume that the tangential restitution coefficient is $\epsilon_\mathrm{t} = 1$, neglecting any exchange between orbital angular momentum and spin angular momentum; furthermore, we assume a normal restitution coefficient $\epsilon_\mathrm{n} = 0.1$, independent of collision velocity. Takeda \& Ida (2001) showed that angular momentum transfer rates are almost independent of $\epsilon_\mathrm{n}$ except for highly elastic cases such as $\epsilon_\mathrm{n} \geq 0.6$. The velocity of each particle after a collision is calculated on the basis of the conservation of momentum. To avoid numerical difficulty with a detailed adjustment of collision and rebounding, we follow Richardson (1994).

The initial condition of the disk is the same as that for run 25 in Kokubo et al. (2000): The initial disk mass is $M_\mathrm{disk} = 4.0 M_L$, and the surface density distribution of the disk is modeled by a power-law distribution $\Sigma \propto a^{-3}$, where $a$ is the semimajor axis of disk particles with inner and outer cutoffs. The inner cutoff is the Earth radius $a_\mathrm{in} = R_{\oplus}$, and the outer cutoff is the Roche limit radius $a_\mathrm{out} = R_\mathrm{Roche}$. All particles have a equal-mass $m$ determined by the number of particles $N$ (i.e., $m = M_\mathrm{disk}/N$). The semimajor axises of particles are randomly chosen between the cutoffs to produce the disk's power-law distribution. Orbital eccentricities and inclinations are given by the Rayleigh distribution, root mean square (RMS) eccentricity $\langle e^2 \rangle ^{1/2} = 0.3$, and RMS inclination $\langle i^2 \rangle ^{1/2} = 0.15$. In general, the initial distributions of eccentricities and inclinations do not affect disk evolution because they relax owing to collisional damping on a timescale of approximately one Kepler period. Eccentric anomaly $E$, longitude of ascending node $\Omega$, and argument of pericenter $\omega$ are randomly chosen between 0 to 2$\pi$. Initial positions and velocities of the particles are uniquely determined by the above parameters ($a$, $e$, $i$, $E$, $\Omega$, and $\omega$).

In this work, we vary the number of particles from $10^4$ to $10^7$, in which one particle corresponds to 10 km-sized satellitesimals. To execute a run with an extra-large number of particles, we deploy FDPS (Framework for Developing Particle Simulators; Iwasawa et al. (2016)), which automatically adopts the tree method (Barnes \& Hut 1986) and parallelizes an arbitrary particle-based numerical simulation method. Furthermore, we used a Pezy-SC processor, which contains a large number of coprocessors to perform massively parallelized numerical simulations (see Appendix).

\section{Results}

\subsection{Overview}

We simulate the evolution of a circumterrestrial disk and lunar accretion and investigate how numerical resolution affects these processes. First, we show the overall evolution of the disk: Figures 1-4 show snapshots of the disk projected onto the $x-y$ plane at $t$ = 0, 1, 5, 10, 20, 40 $ T_\mathrm{K}$,  where we scale the time by $T_\mathrm{K}$, which is the Kepler period at the Roche limit distance. Figure 1 in this paper corresponds to Fig. 4 in Kokubo et al. (2000) (i.e., $N = 10^4$). For high-resolution simulations ($N \geq 10^6$), snapshots at $t$ = 60, 80, 100, 110 $T_\mathrm{K}$ are also shown in Figs. 5 and 6. The point sizes in Figures 1-6 are proportional to the physical size of disk particles.

\begin{figure*}
\plotone{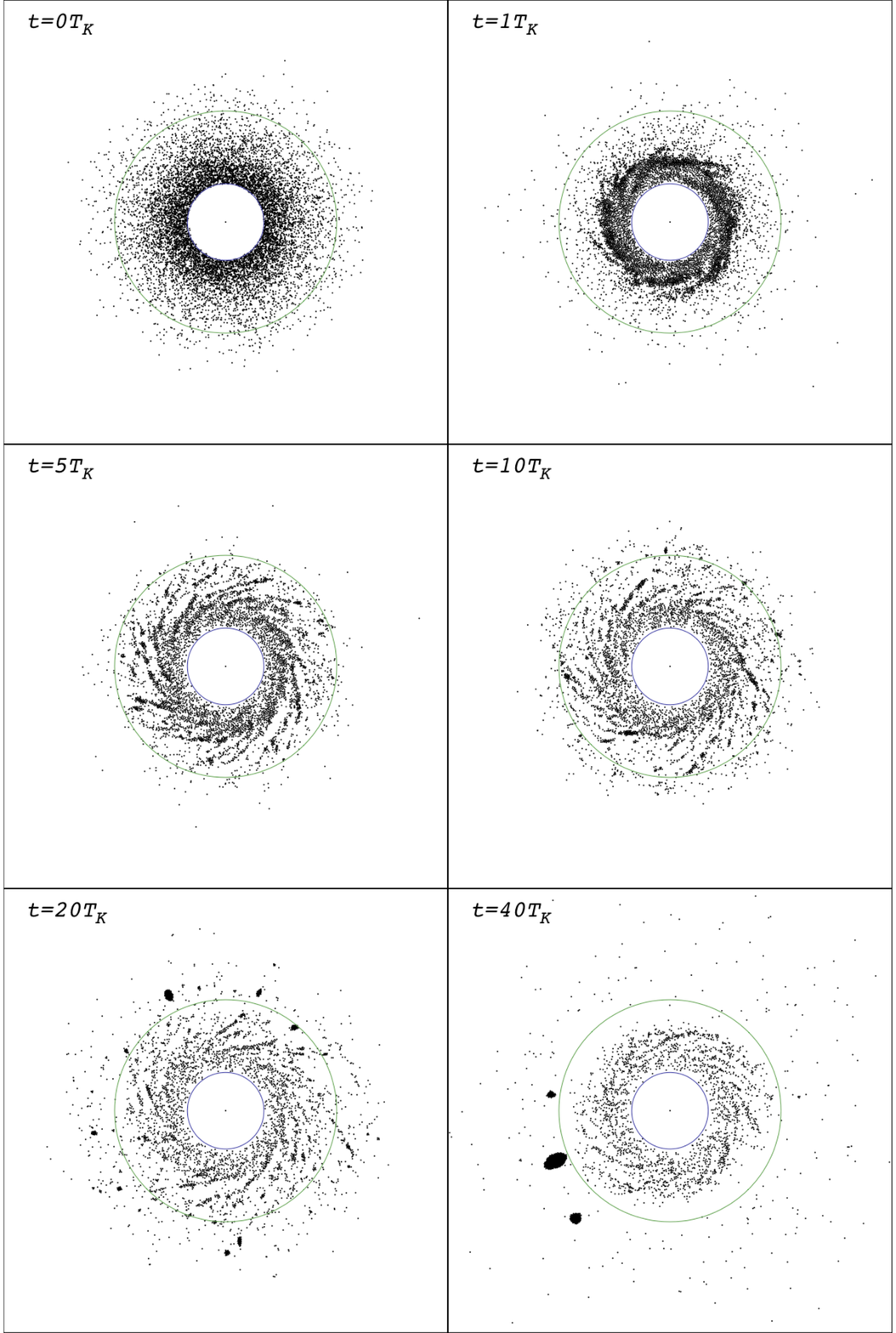}
\caption{Snapshots of circumterrestrial disk projected onto the $x$-$y$ plane at $t$ = 0, 1, 5, 10, 20, 40 $T_\mathrm{K}$. The solid blue circle centered on the coordinate origin represents Earth, and the green circle is the Roche limit radius. The number of particles is $10^4$.}
\end{figure*}

\begin{figure*}
\plotone{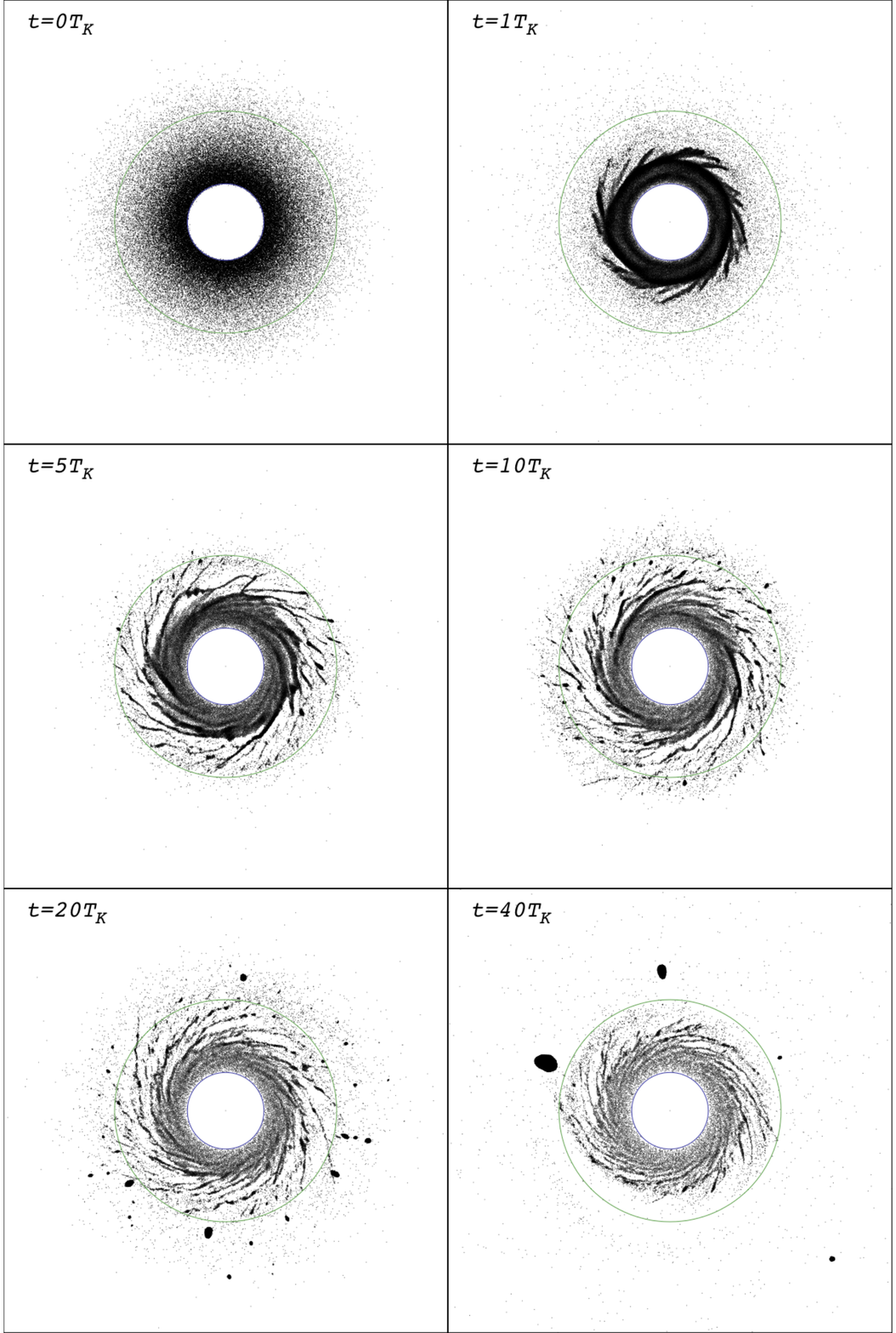}
\caption{The same snapshots as in Fig. 1 but the number of particles is $10^5$.}
\end{figure*}

\begin{figure*}
\plotone{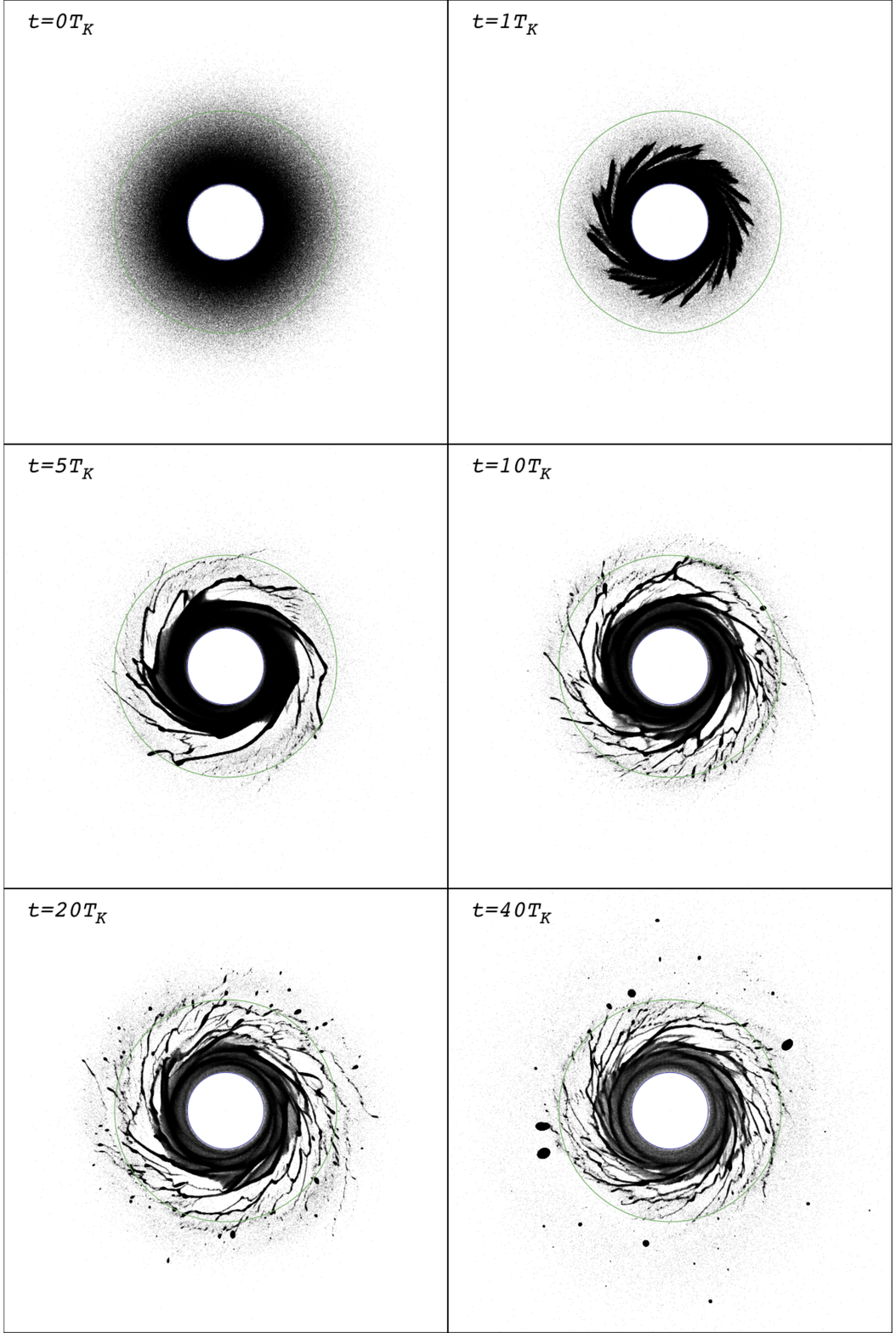}
\caption{The same snapshots as in Fig. 1 but the number of particles is $10^6$.}
\end{figure*}

\begin{figure*}
\plotone{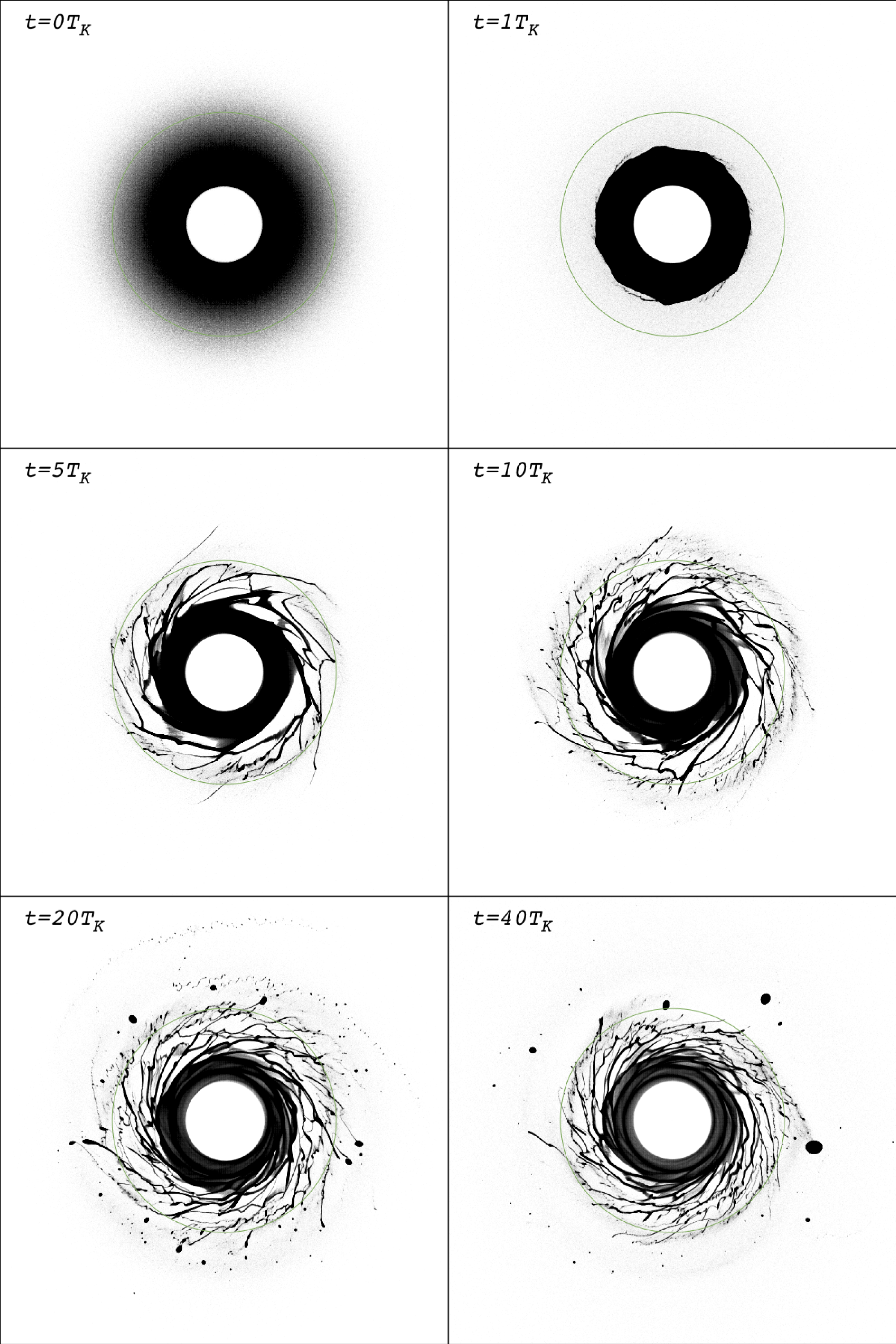}
\caption{The same snapshots as in Fig. 1 but the number of particles is $10^7$.}
\end{figure*}

\begin{figure*}
\plotone{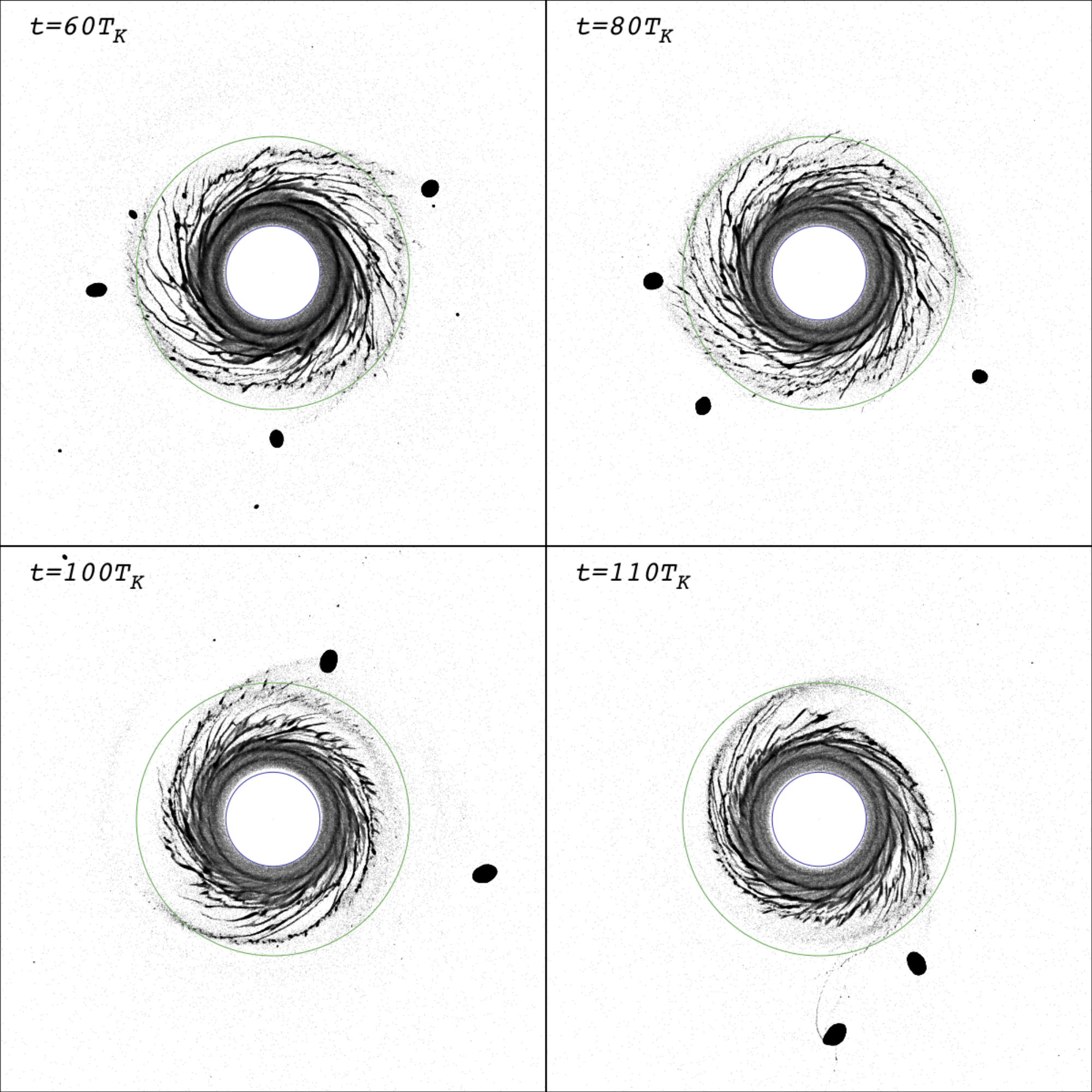}
\caption{The same snapshots as in Fig. 1 but at $t$ = 60, 80, 100, 110 $T_\mathrm{K}$.}
\end{figure*}

\begin{figure*}
\plotone{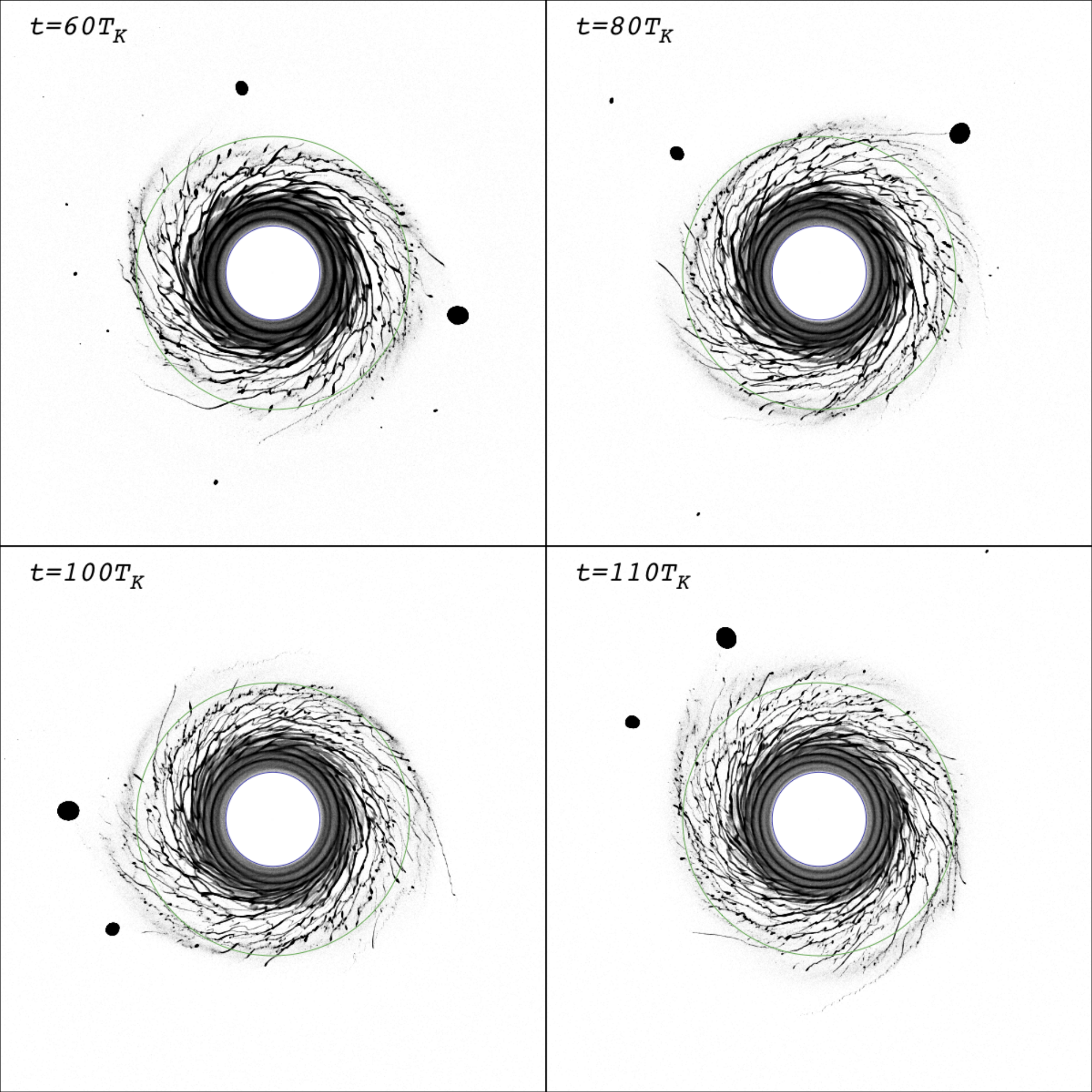}
\caption{The same snapshots as in Fig. 5 but the number of particles is $10^7$.}
\end{figure*}

The evolution of the circumterrestrial disk for the simulation with $N = 10^4$ (Fig. 1) is consistent with that found by Kokubo et al. (2000). The disk contracts by the collisional damping of particles, and particle clumps grow inside the Roche limit radius ($t=T_\mathrm{K}$). The clumps are elongated by Keplerian shear and spiral arms are formed ($t=5T_\mathrm{K}$). Particles are transferred beyond the Roche limit radius via the gravitational torque exerted by the spiral arms ($t=10T_\mathrm{K}$). Small particles beyond the Roche limit radius collapse to form small moonlets, the rapid accretion of which forms a lunar seed ($t=20T_\mathrm{K}$). The seed grows exclusively by sweeping up particles transferred beyond the Roche limit radius, and the proto-Moon forms just outside the Roche limit radius ($t=40T_\mathrm{K}$). Although relatively large moonlets survive on the horseshoe orbit with the proto-Moon, they collide with each other over the long-term tidal interaction with Earth to accrete to a single moon.

\subsection{Mass Evolution of the Proto-Moon}

Figures 1-4 show the evolution of the disks, which are qualitatively the same as that of Kokubo et al. (2000) regardless of the numerical resolution. However, the spiral arm structure differs between low-resolution simulations ($N \leq 10^5$) and high-resolution simulations ($N \geq 10^6$), and this difference results in a longer timescale for lunar accretion in the latter case. For low-resolution simulations, individual spiral arms appear clearly, and the proto-Moon is almost formed at $t=40T_\mathrm{K}$. Conversely, for high-resolution simulations, the spiral arms connect with each other to form more complicated structures between the arms, and the proto-Moon forms at approximately $t=100 T_\mathrm{K}$.

The mass evolution of the largest aggregate and that of the sum of the three largest aggregates are plotted as a function of time in Fig. 7. These graphs clearly show that the mass evolution processes for low-resolution simulations ($N \leq 10^5$) differ from those of high-resolution simulations ($N \geq 10^6$). The low-resolution simulations show that lunar accretion is divided into two stages: rapid growth ($t \leq 40T_\mathrm{K}$) and saturated growth ($t > 40 T_\mathrm{K}$). The proto-Moon is almost formed in the first stage. In the second stage, the proto-Moon not only terminates its growth but also decreases its mass because of the collisional and/or tidal strip caused by gravitational interactions with disk material and/or other moonlets. Conversely, high-resolution simulations show that the lunar seed grows continuously and gradually until the final mass is reached at $t \sim 100 T_\mathrm{K}$.

\begin{figure*}
\plotone{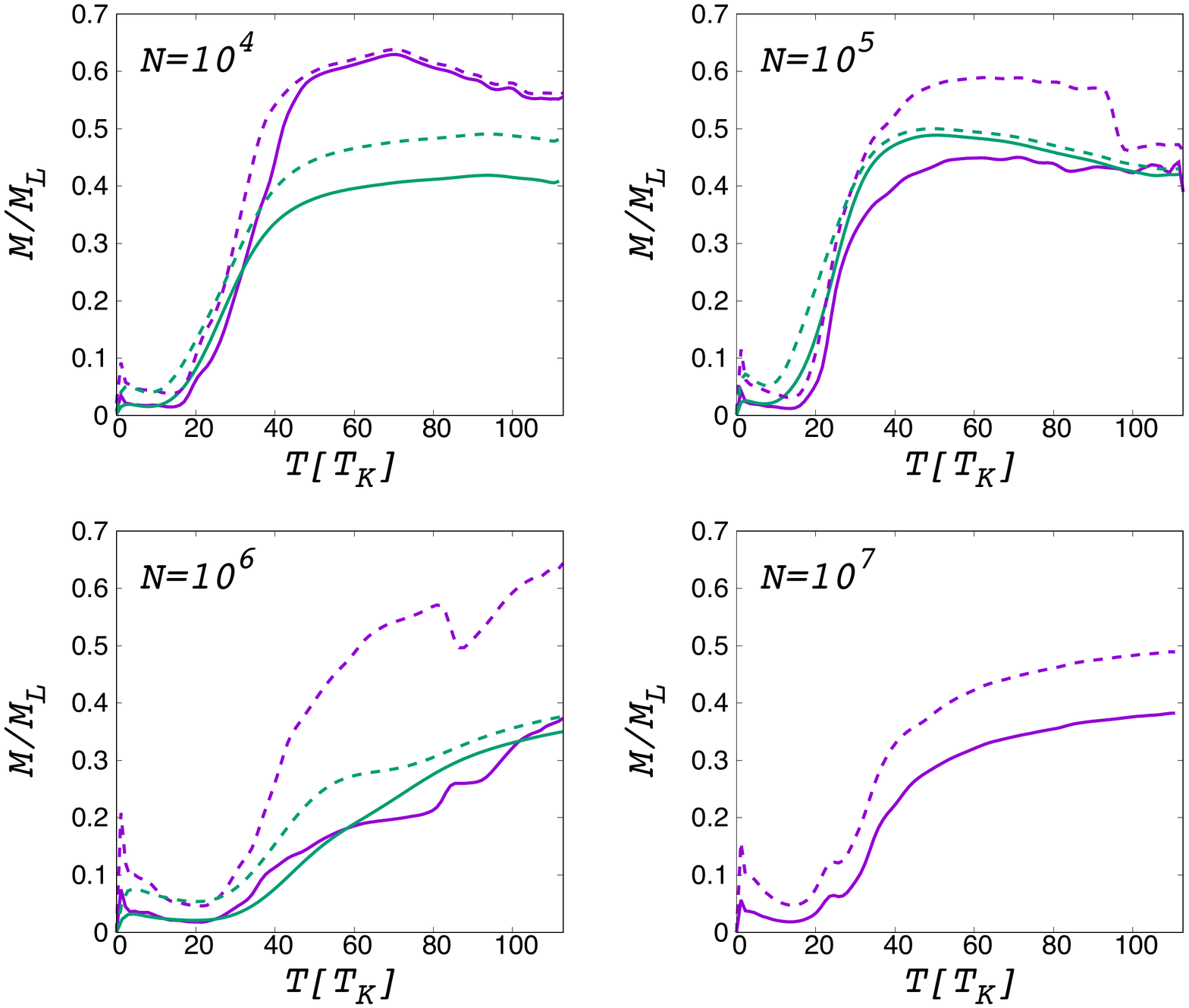}
\caption{Mass evolution for the largest aggregate (lines) and for the sum of three largest aggregates (dotted lines) normalized by lunar mass $M_L$ for $N=10^4$, $10^5$, $10^6$, and $10^7$. Each panel except for $N=10^7$ shows two sets of results with the same initial conditions but different random seeds to generate the initial particle positions and velocities.}
\end{figure*}

On the other hand, the accumulative masses accreted onto the Earth are plotted as a function of time in Fig. 8. The mass of over 2 $M_L$, which corresponds to a half of the initial disk mass, has fallen to the Earth in 40 $T_K$. The accreted masses for low-resolution simulations ($N \leq 10^5$) are slightly larger than those for high-resolution simulations ($N \geq 10^6$). However, in contrast to the mass evolution of the proto-Moon, the convergent behavior of accreted mass does not depend on the numerical resolution.

\begin{figure}
\plotone{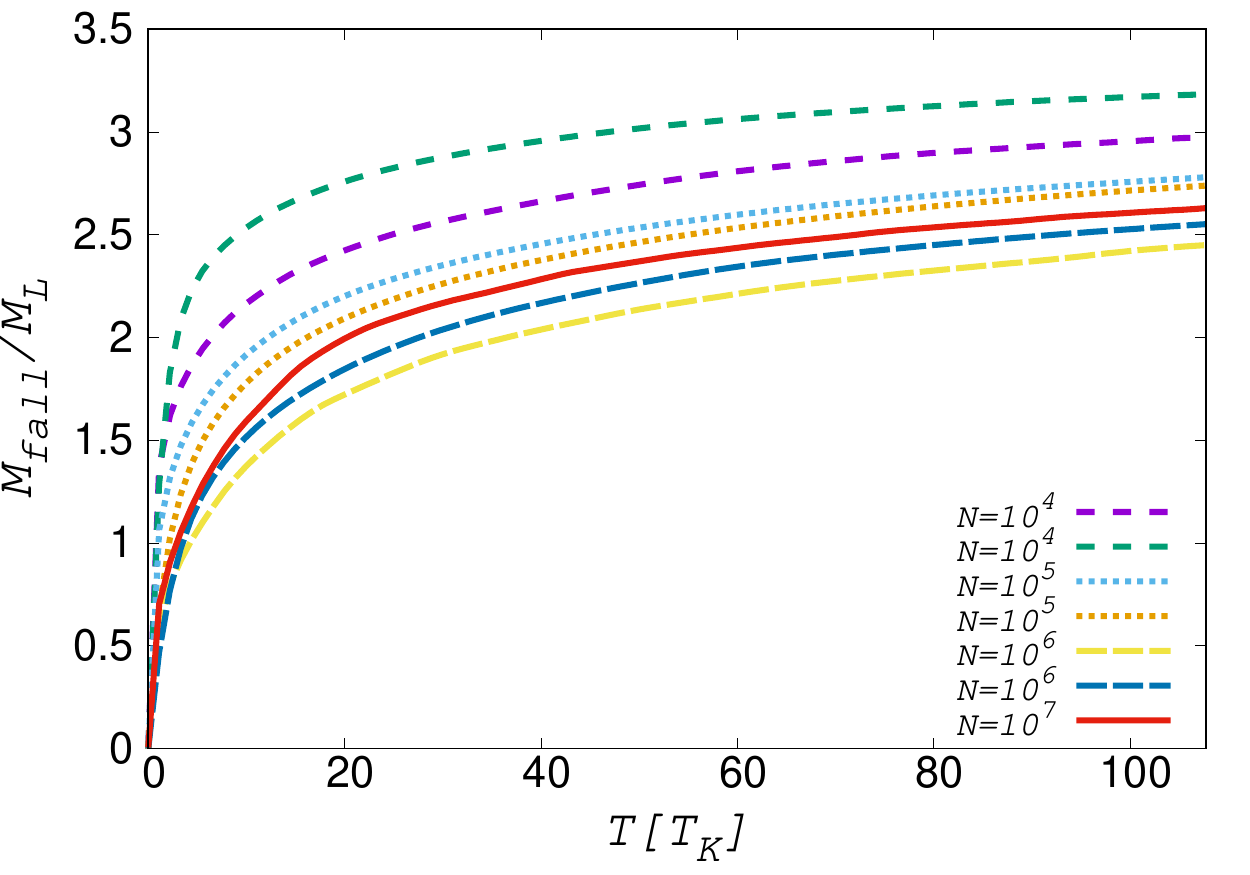}
\caption{Accumulative masses accreted onto the Earth for $N=10^4$, $10^5$, $10^6$, and $10^7$. Each panel except for $N=10^7$ shows two sets of results with the same initial conditions but different random seeds to generate the initial particle positions and velocities.}
\end{figure}

The simulated mass of the largest aggregate, that of the sum of the three largest aggregates, and that fallen to the Earth at $t=110T_\mathrm{K}$ are summarized in Table 1 and Fig. 9. Figure 9 indicates that both the Moon's mass and the accumulative mass fallen to the Earth would not depend on the numerical resolution. The Moon's masses are smaller than those reported by Kokubo et al. (2000) with the same disk parameters but different accretion models (i.e., total and partial accretion models). Considering that we can use the rubble pile model to simulate processes such as collisional breakup and/or tidal strip for a proto-Moon, the final mass of the proto-Moon (Table 1) would be smaller than that obtained by Kokubo et al. (2000) (Table 2). Although stripping the proto-Moon's material would not qualitatively change the mode of lunar accretion, it can quantitatively change the final lunar mass. Forming a Moon body requires a longer time because of the re-accretion of the stripped materials to the proto-Moon and/or a more massive circumterrestrial disk made by the giant impact. However, note that aggregates are too fragile to withstand collisional breakup and tidal strip in the rubble pile model because the model neglects the strength of the materials. Therefore, the actual masses of the proto-Moon produced in the disk would be slightly greater than those listed in Table 1.

\begin{deluxetable}{c|ccc}
\tablecaption{Final mass of the proto-Moon and the mass fallen to the Earth}
\tablehead{
\colhead{$N$} & \colhead{$M[M_L]$} & \colhead{$M_3[M_L]$} & \colhead{$M_{\mathrm fall}[M_L]$}
}
\startdata
$10^4$ & 0.556 & 0.562 & 2.980\\
$10^4$ & 0.409 & 0.481 & 3.186\\
$10^5$ & 0.390 & 0.462 & 2.786\\
$10^5$ & 0.420 & 0.430 & 2.742\\
$10^6$ & 0.374 & 0.644 & 2.457\\
$10^6$ & 0.352 & 0.394 & 2.554\\
$10^7$ & 0.383 & 0.490 & 2.636\\
\enddata
\tablecomments{The mass of the largest aggregate $M$, the mass of the sum of the three largest aggregates $M_3$, and the mass fallen to the Earth $M_{\mathrm fall}$ at $t=110 T_\mathrm{K}$ for $N=10^4$, $10^5$, $10^6$, $10^7$. The masses are normalized by the present lunar mass $M_L = 0.0123 M_{\oplus}$. Each row except for $N=10^7$ shows two sets of the results with the same initial conditions but different random seeds to generate the initial particle positions and velocities.}
\end{deluxetable}

\begin{figure}
\plotone{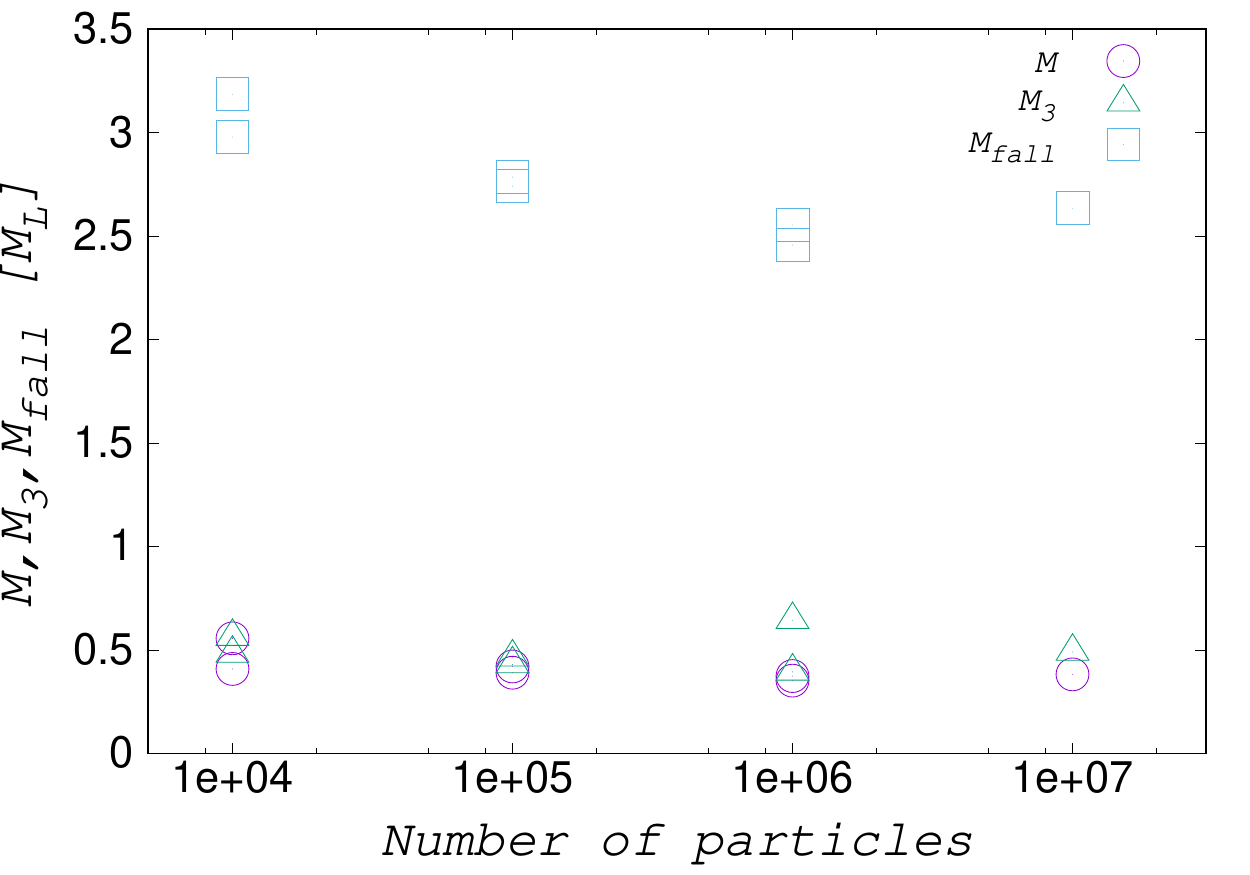}
\caption{Plot of the data in Table 1 to see the convergent behavior with numerical resolution.}
\end{figure}

\begin{deluxetable}{c|cc}
\tablecaption{Final mass of the proto-Moon found by \citet{kokubo00}}
\tablehead{
\colhead{$Run$} & \colhead{$M_{f=0.7}[M_L]$} & \colhead{$M_{f=1.0}[M_L]$}
}
\startdata
$13$ & 0.925 & 0.834 \\
$25$ & 0.875 & 0.676 \\
$26$ & 0.957 & 0.965 \\
$27$ & 0.640 & 0.872 \\
$30$ & 0.895 & 0.860 \\
\enddata
\tablecomments{The mass of the largest aggregate for the partial ($f=0.7$) and total ($f=1.0$) accretion models in Kokubo et al. (2000). We only show limited results from runs in which the initial conditions are the same as those in our simulations. The masses are normalized by the present lunar mass $M_L$.}
\end{deluxetable}

\subsection{Disk Evolution}
We now discuss in detail the evolution of the circumterrestrial disks. Figures 10, 11 and 12 show snapshots of the surface density $\Sigma$, the radial component $v_R$ of the velocity dispersion, and Toomre's $Q$ value, respectively. The velocity dispersion is the RMS deviation velocity from the local Keplerian circular velocity. Toomre's $Q$ value for a particle disk is approximately
\begin{equation}
Q \equiv \frac{v_{\mathrm R} \Omega}{\pi G \Sigma},
\end{equation}
where $\Omega = 2\pi / T_\mathrm{K}$ is the Keplerian angular frequency (Toomre 1964). A linear stability analysis of a self-gravitating disk shows that the disk is stable against axisymmetric perturbation when $Q > 1$. The figures show that the evolution of each parameter is consistent with the results of Kokubo et al. (2000) regardless of the numerical resolution. Therefore, the disk evolution process is essentially independent of numerical resolution.

\begin{figure*}
\plotone{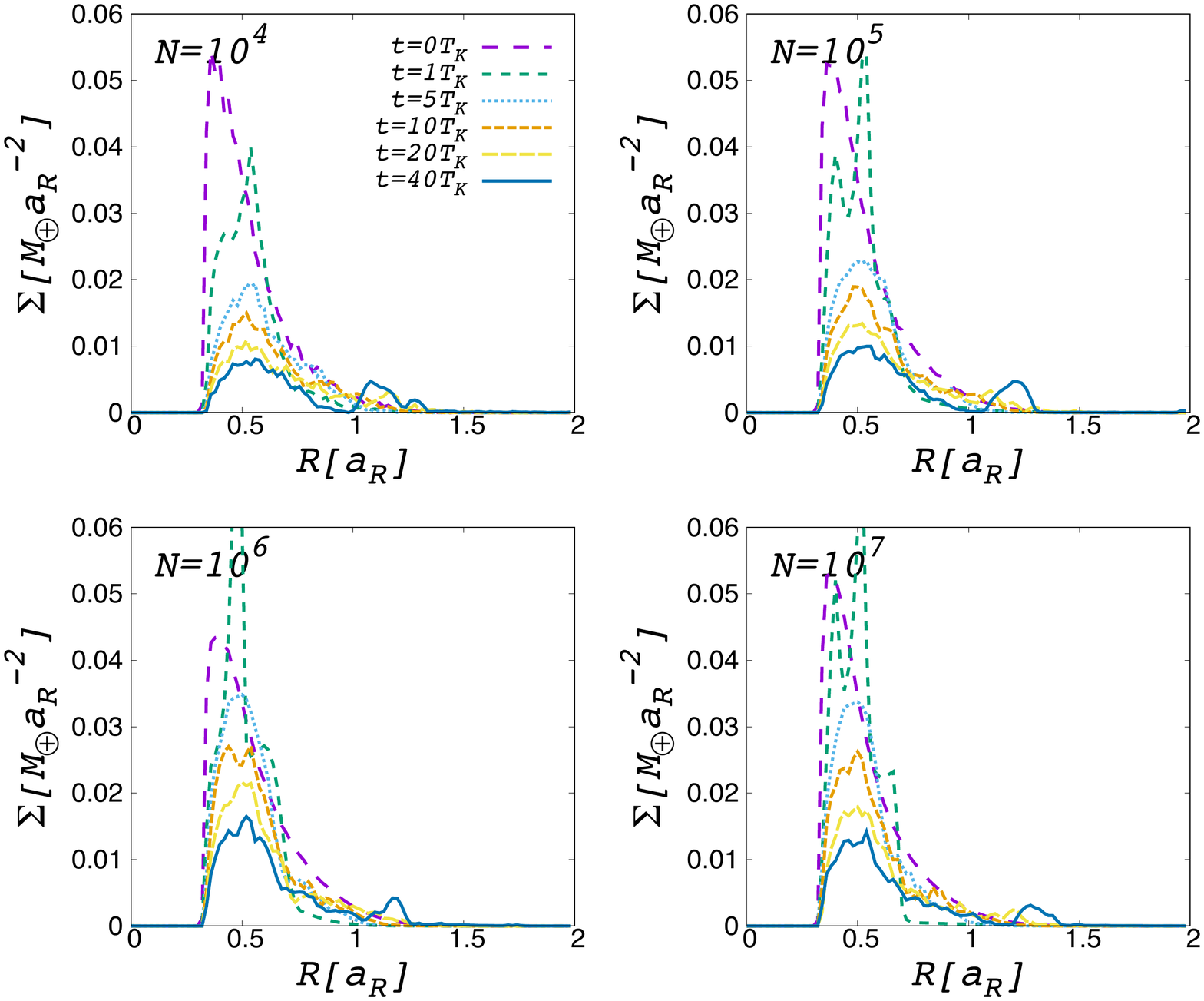}
\caption{Surface density $\Sigma$ plotted as a function of distance from Earth for $N=10^4$, $10^5$, $10^6$, and $10^7$ at $t$ = 0, 1, 5, 10, 20, 40 $ T_\mathrm{K}$.}
\end{figure*}

\begin{figure*}
\plotone{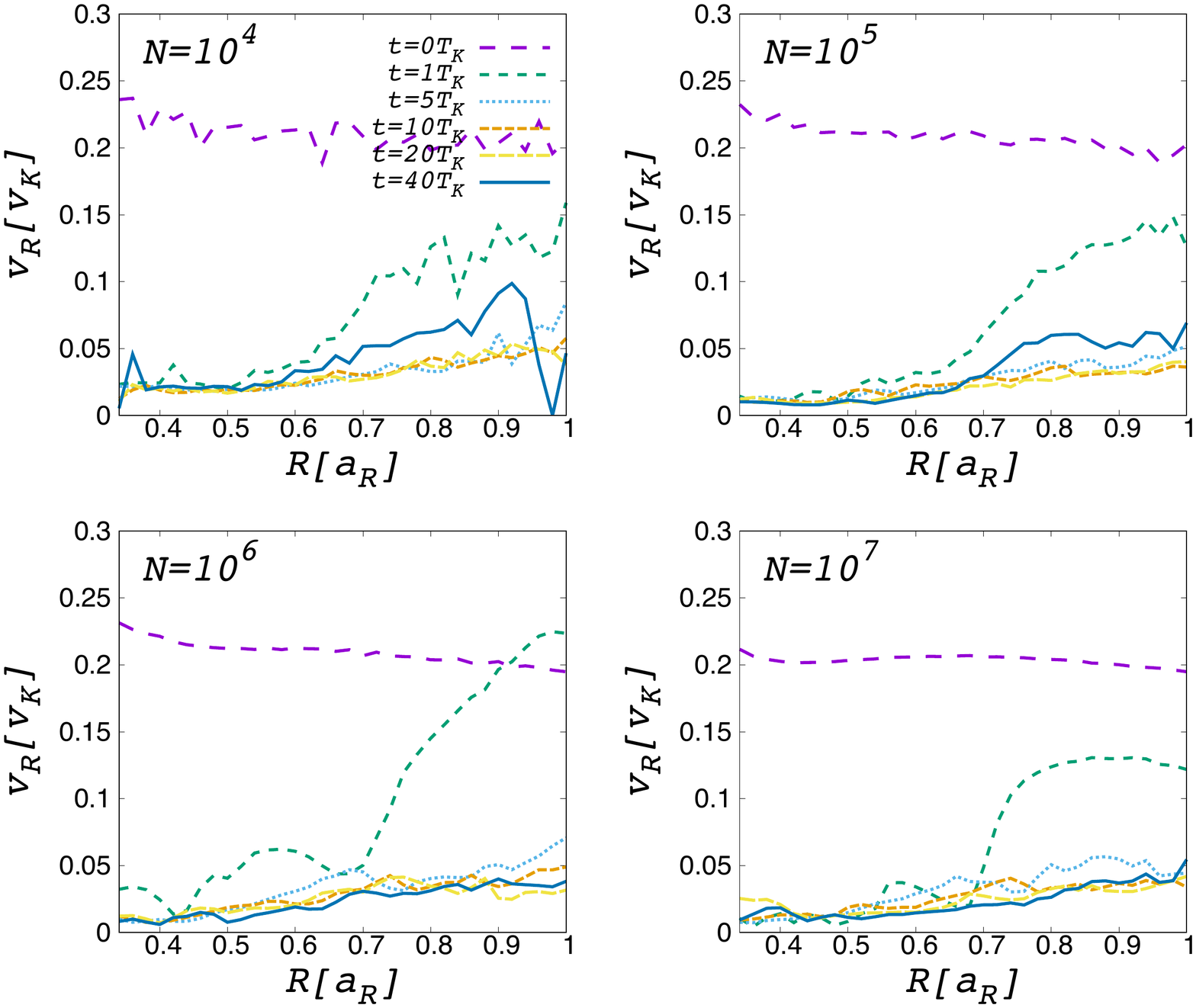}
\caption{Radial profile of RMS velocity dispersion in the radial direction of disk particles for $N=10^4$, $10^5$, $10^6$, and $10^7$ at $t$ = 0, 1, 5, 10, 20, 40 $ T_\mathrm{K}$ (the legend is the same as for Fig. 10).}
\end{figure*}

\begin{figure*}
\plotone{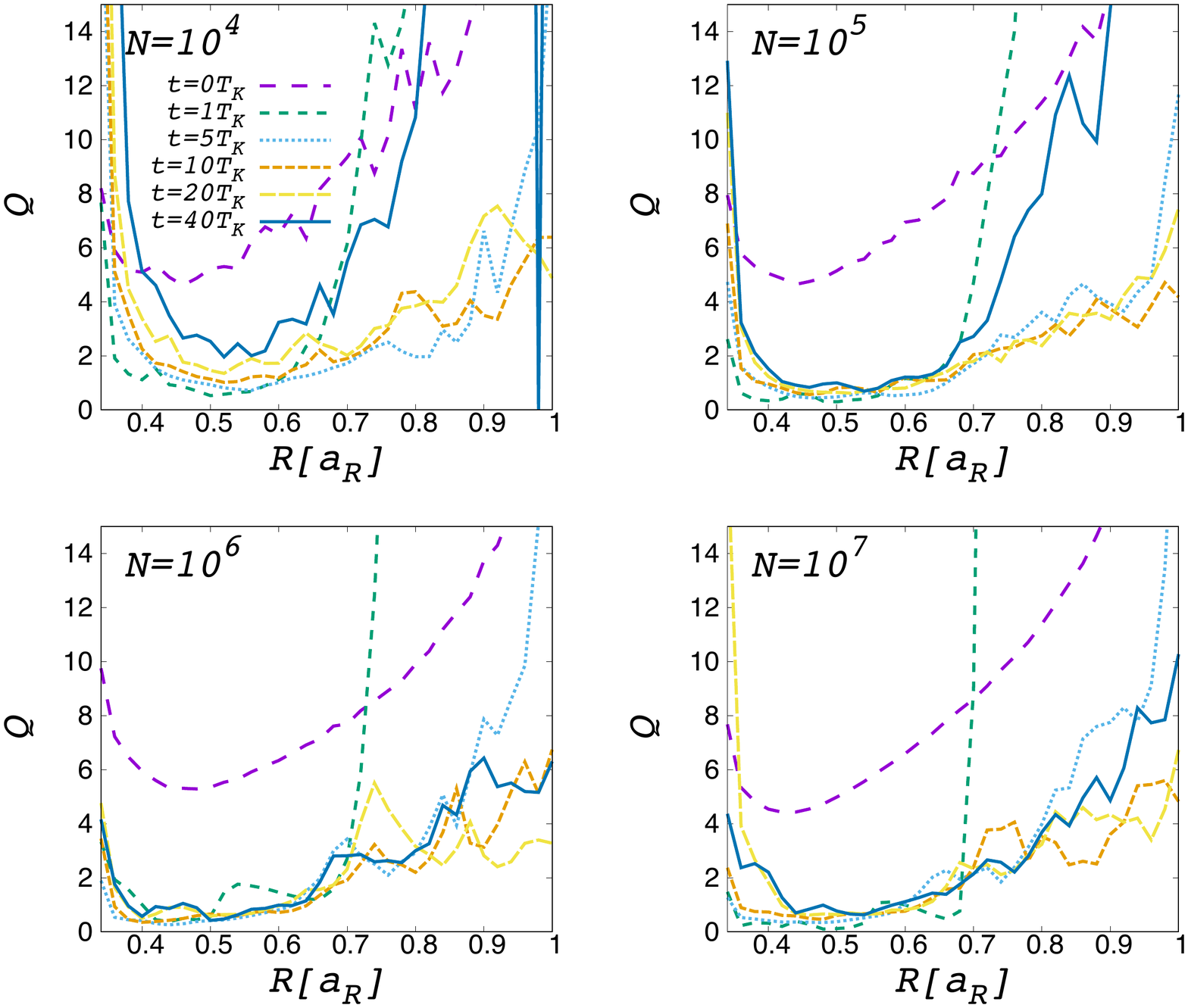}
\caption{Toomre's $Q$ value as a function of distance from Earth for $N=10^4$, $10^5$, $10^6$, and $10^7$ at $t$ = 0, 1, 5, 10, 20, 40 $ T_\mathrm{K}$ (the legend is the same as for Fig. 10).}
\end{figure*}

When $Q$ has its minimum value ($Q \simeq 1$), the circumterrestrial disk becomes marginally gravitationally unstable, and the critical wavelength expected from the linear stability analysis (see, e.g., Toomre 1964) is
\begin{equation}
\lambda_{\mathrm c} \equiv \frac{2\pi^2G\Sigma}{\Omega^2} \sim 2\pi R \frac{M_{\mathrm {disk}}}{M_{\oplus}}.
\end{equation}
Given the moderate pitch angle of the spiral arms, the number of spiral arms may be estimated by
\begin{equation}
n_{\mathrm s} \simeq \frac{2\pi R}{\lambda_c} \sim \frac{M_{\oplus}}{M_{\mathrm {disk}}}.
\end{equation}
Our simulation ($M_{\mathrm {disk}} = 4 M_{\mathrm L}$) corresponds to $n_{\mathrm s} \sim 20$. Considering that the Moon is primarily formed by the material transferred beyond the Roche limit radius, the timescale of Moon formation is almost equivalent to the timescale of the mass or angular momentum transfer in the disk due to the gravitational torque exerted by the spiral arms.

We estimate the number of spiral arms for each simulation using the mode $m$ that has the maximum amplitude of the Fourier coefficient of the azimuthally averaged surface density of the disk. (Please see Michikoshi \& Kokubo (submitted) for details.) Figure 13 shows that the number of spiral arms differs between low-resolution simulations ($N \la 10^5$) and high-resolution simulations ($N \ga 10^6$). In the former case, the spiral arms number are around 20, which is consistent with the above analytical estimate and with the numerical results of Kokubo et al. (2000). Conversely, in the latter case, the spiral arms always number less than 10, and no spiral arms are recognized for $t > 80 T_\mathrm{K}$ and $N=10^7$.

\begin{figure}
\plotone{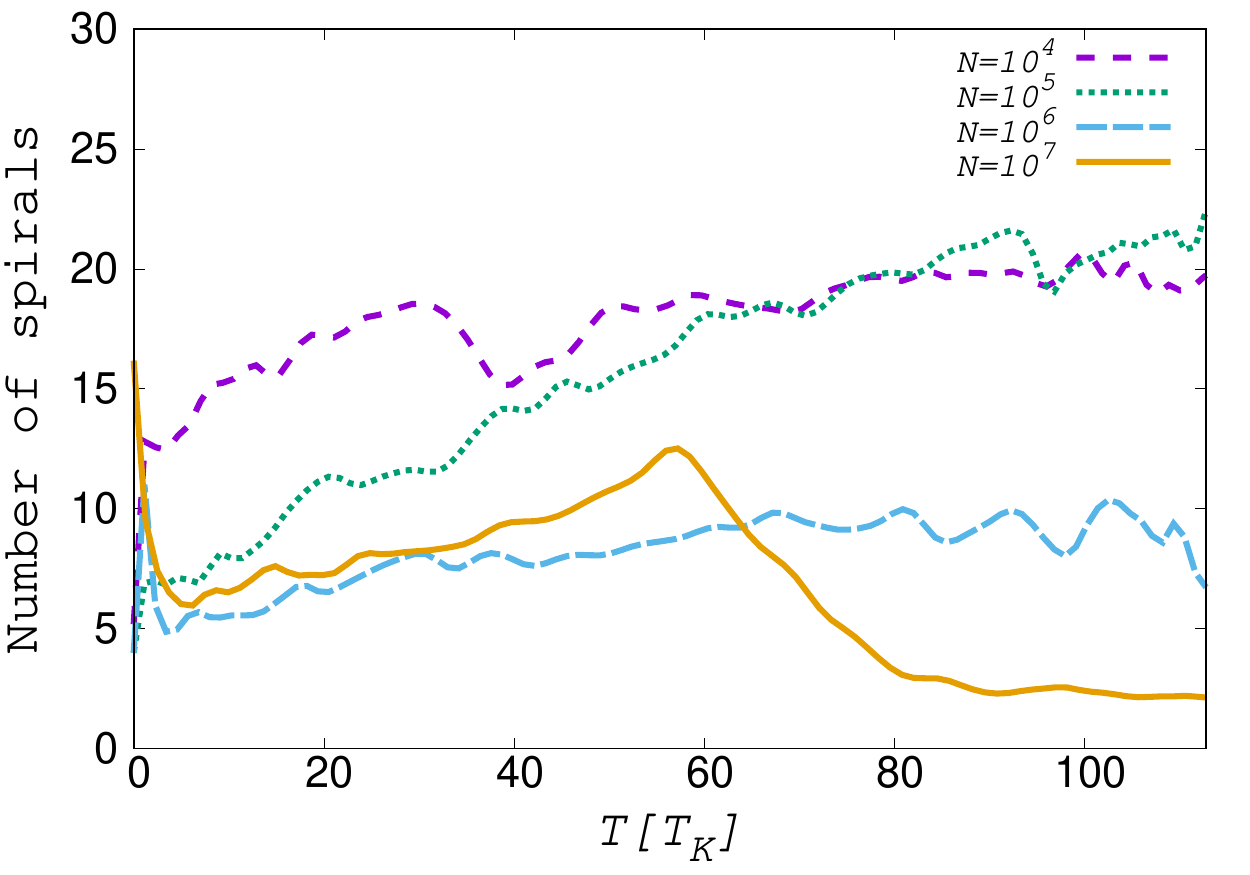}
\caption{Evolution of number of spiral arms for $N=10^4$, $10^5$, $10^6$, and $10^7$. The analysis code was provided by S. Michikoshi.}
\end{figure}

\section{Summary and Discussions}

We executed $N$-body simulations of varying numerical resolution ($N=10^4-10^7$) of lunar accretion from a circumterrestrial disk. Irrespective of the numerical resolution, the proto-Moon forms just outside the Roche limit radius. We find that the spiral structure inside the Roche limit radius differs between low-resolution simulations ($N \leq 10^5$) and high-resolution simulations ($N \geq 10^6$). According to the difference, the angular momentum fluxes determined by the spiral arms and those determined by the accretion timescales of the Moon also depend on the numerical resolution.

For high-resolution simulations, the spiral arms are connected to each other, forming more complicated structures between the spiral arms. The complicated structures resemble the gravitational wake structures that appear in the numerical simulations for a dense planetary ring (Daisaka et al. 2001; Michikoshi \& Kokubo 2011). The self-gravity wakes are small nonaxisymmetric spiral structures formed by gravitational instability. The wake structure is time dependent and transient, being created and destroyed on a timescale of the order of a Keplerian period. Daisaka \& Ida (1999) showed that as the wake structure grows, the radial velocity dispersion increases with a large magnitude of fluctuation and tends to maintain the relationship $Q \sim 2$. Daisaka et al. (2001) showed that when the wake structure is developed by gravitational instability, angular momentum transport is significantly enhanced not only by gravitational torque due to the wake-like structure but also by the associated coherent motion of the wakes. Takeda \& Ida (2001) also showed that the angular momentum transferred by nonaxisymmetric clumps is always as large as that due to the gravitational torque exerted by the spirals. Therefore, the self-gravity wakes cause efficient radial diffusion in the circumplanetary disk. However, our high-resolution simulations indicate that angular momentum transfer becomes inefficient when the wake-like structures appear. Therefore, we cannot conclude that the complicated structures that appear in the high-resolution simulations are self-gravity wakes. 

Another candidate for this complicated structure is a coexistence of nonaxisymmetric spiral arms and axisymmetric structures. The axisymmetric structure is caused by overstable oscillations (Schmit \& Tscharnuter 1995). The viscous overstability appears in very-high-resolution simulations and can be a good candidate to explain the small structures in the planetary rings (Daisaka et al. 2001). Such overstable oscillations would play an important role in the angular momentum transport in the circumterrestrial disk and significantly modify the angular momentum flux. However, although Salo (1992) advanced the study of the self-gravity and viscous overstability mechanisms through numerical $N$-body simulations, no analytical theory can currently model the formation of overstability with self-gravity spirals or wakes (Ballouz et a. 2017). Therefore, a comprehensive explanation of the interplay between the mechanisms of self-gravity and viscous overstability remains elusive.

In this paper, we focus on evaluating how the lunar accretion process depends on the number $N$ of particles in an $N$-body simulation and find qualitative and quantitative differences in the Moon formation process. In future works, we plan to execute a large number of high-resolution $N$-body simulations of Moon formation with a variety of disk parameters to investigate the physics that would cause such differences.

\acknowledgments

We acknowledge the reviewer's helpful comments. We thank Shugo Michikoshi for providing an analysis code and valuable discussions. We also thank the people at Pezy Computing for their help in coding on Pezy-SC. We used the super computer Shoubu at RIKEN. This work was supported by JSPS KAKENHI Grant Number JP15K17750 and MEXT Grant Number 26106006.

\appendix

\section{Pezy-SC Processors}

The Pezy-SC processor is a novel new architecture that integrates 1024-core MIMD processors with a physically shared memory and hierarchical cache. It was developed by the Japanese venture company Pezy Computing. The Pezy-SC processor has achieved impressive performance per watt: The ExaScaler system based on the first-generation Pezy-SC ranked second in the Green500 list in November 2014, and three ExaScaler systems occupied the top three ranks in the Green500 list in June 2015. Please see Yoshifuji et al. (2016) for details on the Pezy-SC processor.

In order to show the speed-up efficiency of the PEZY-SC device, we briefly show the measured wall-clock times of our code by changing the number of particles deployed in a run in Figure 14. For a comparison, we also show the wall-clock time measured on a CPU (Intel Xeon E5-2618L v3). The wall-clock time of our code becomes roughly $N \log_8 N$. The speed-up efficiency of PEZY-SC compared to CPU reaches to about $20$ times faster. In the case of $N = 10^7$, the wall-clock time with the number of processes of $32$ is about $9$ seconds. In our simulation, the end time is set to $3500$ and the timestep is $2^{-9}$. In this case, if the number of particles is constant, the estimated wall-clock time to finish the run is $3500 \times 512 \times 9$ seconds which corresponds to about half a year. Recall that the number of particles in one run will decrease to a third of the initial number of particles. Thus, the actual wall-clock time is less than the estimated value. In the case, the actual value is about 3 months. Note that our code surely has room for further speed-ups. Since the main purpose of this paper is the Moon formation, detailed measurement and breakdowns (e.g., wall-clock time on the host v.s. the device and strong/weak scaling to the number of devices) should be given in another context.

\begin{figure}
\plotone{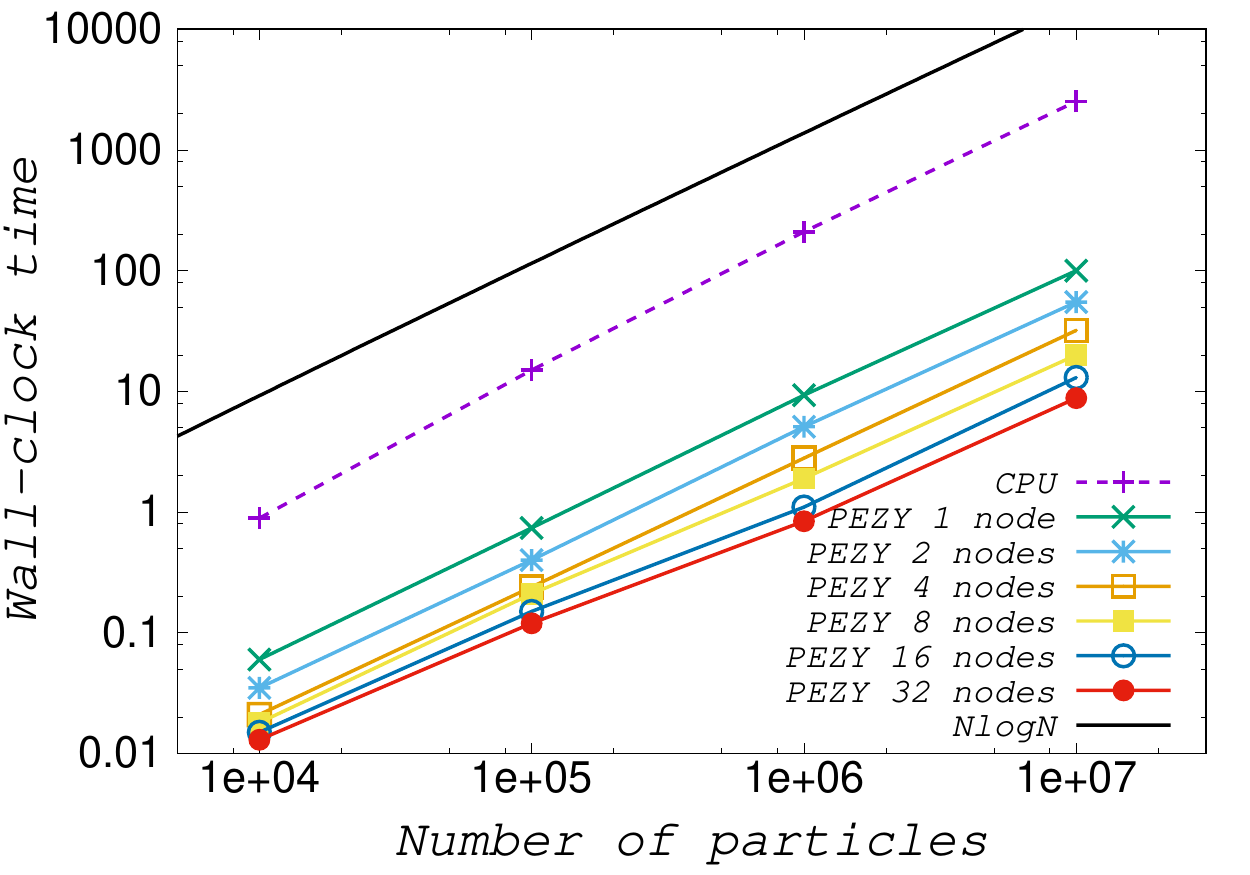}
\caption{Measured wall-clock time of our code on a CPU and PEZY-SC for $N=10^4$, $10^5$, $10^6$, and $10^7$ with a line of $N \log_8 N$.}
\end{figure}

\end{document}